\documentstyle[amssymb,aps]{revtex}

\input epsf

\begin{document}
\title{Giant dispersion of critical currents in superconductor with fractal
clusters of a normal phase}
\author{Yuriy I. Kuzmin}
\address{Ioffe Physical Technical Institute of the Russian Academy of Sciences,\\
Polytechnicheskaya 26 St., Saint Petersburg 194021 Russia,\\
and State Electrotechnical University of Saint Petersburg,\\
Professor Popov 5 St., Saint Petersburg 197376 Russia\\
e-mail: yurk@mail.ioffe.ru; iourk@yandex.ru\\
tel.: +7 812 2479902; fax: +7 812 2471017}
\date{\today}
\maketitle
\pacs{74.60.Ge; 74.60.Jg}

\begin{abstract}
The influence of fractal clusters of a normal phase on the
dynamics of a magnetic flux trapped in a percolative
superconductor is considered. The critical current distribution
and the current--voltage characteristics of fractal
superconducting structures in the resistive state are obtained for
an arbitrary fractal dimension of the cluster boundaries. The
range of fractal dimensions, in which the dispersion of critical
currents becomes infinite, is found. It is revealed that the
fractality of clusters depresses of the electric field caused by
the magnetic flux motion thus increasing the critical current
value. It is expected that the maximum current-carrying capability
of a superconductor can be achieved in the region of giant
dispersion of critical currents.
\end{abstract}

\bigskip

One way to increase the critical currents of superconductors
consists in creating the artificial pinning centers in the volume
of material. The role of such centers can be performed by the
clusters of a normal phase formed in the course of the film growth
at the sites of defects at the film-substrate interface
\cite{mez}--\cite{pss}. New possibilities for the pinning
enhancement are opened in the case of clusters of a normal phase
with
fractal boundaries \cite{sur}--\cite{prb} (see also references in \cite{prb}%
). In the present paper the effect of such fractal clusters on the critical
currents and current-voltage characteristics of superconductors in the
resistive state are considered.

The problem setting is similar to that considered in \cite{pla},
\cite{prb}. A superconductor containing columnar inclusions of a
normal phase is cooled in a magnetic field below the critical
temperature (in field-cooling regime). As a result, the magnetic
flux is trapped in isolated clusters of a normal phase. Then a
transport current is passed through the sample across the
direction of the magnetic field. It is assumed that a
superconducting percolation cluster has been formed in the plane
where electric current flows. While the transport current
increases, the trapped flux remains unchanged until the vortices
begin to break away from the clusters for which the pinning force
is smaller than the Lorentz force created by the current. The
vortices travel through the superconducting space via weak links
between clusters of a normal phase. The weak links are readily
formed at various structural defects in high temperature
superconductors (HTS's) characterized by a small coherence length.
In usual low-temperature superconductors the weak links can be
formed as a result of the proximity effect at the sites of minimum
distance between the next normal phase clusters.

Thus, irrespective of their nature, the weak links form channels
for the transport of vortices. According to the weak link
configuration, each cluster of the normal phase contributes to the
total distribution of the depinning critical currents. A geometric
probability analysis of the influence of distribution of the
points, where vortices can enter into weak links, on the critical
currents of clusters was performed in Ref.~\cite{prb}. As the
transport current increases, the vortices break away first from
the clusters of a smaller pinning force and, accordingly, a
smaller critical current. Therefore, a change in the trapped
magnetic flux $\Delta \Phi $ is proportional to the number of
normal phase clusters of the critical currents below a preset
value of $I$. Therefore, a relative flux decrease is equal to the
probability that the critical current $I^{\prime }$ of an
arbitrarily selected cluster is smaller than $I$:
\begin{equation}
\frac{\Delta \Phi }{\Phi }=\int\limits_{0}^{I}f\left( I^{\prime }\right)
dI^{\prime }  \label{eq1}
\end{equation}
where $f\left( I\right) $ is the probability density for the distribution of
the critical depinning currents.

At the same time, the magnetic flux trapped in a cluster is
proportional to the cluster area $A$. So, a change in the total
trapped flux depends on the distribution of areas of the normal
phase clusters. Thus, in order to determine how the transport
current affects the trapped magnetic flux, it is necessary to find
a relationship between distributions of the critical currents of
clusters and their areas. This problem was solved in Ref.~\cite
{pla2} for the general case of the gamma-distribution of areas of
the clusters with fractal boundaries:
\begin{equation}
w\left( a\right) =\frac{\left( g+1\right) ^{g+1}}{\Gamma \left( g+1\right) }%
a^{g}\exp \left( -\left( g+1\right) a\right)  \label{eq2}
\end{equation}
where $w(a)$ is the probability density for the cluster area distribution,
$a\equiv A/\overline{%
A}$ is the dimensionless area of a cluster, $\overline{A}$ is the
average cluster area, $g$ is the gamma-distribution parameter
which determins the
standard deviation of the cluster area $\sigma _{a}=1/\sqrt{g+1}$, and $%
\Gamma (\nu)$ is the Euler gamma--function. In this case, the
distribution of critical currents is as follows:
\begin{equation}
f(i)=\frac{2G^{g+1}}{D\Gamma (g+1)}i^{-\left( 2/D\right) (g+1)-1}\exp \left(
-Gi^{-2/D}\right)  \label{eq3}
\end{equation}
where
\[
G\equiv \left( \frac{\theta ^{\theta }}{\theta ^{g+1}-\left( D/2\right) \exp
\left( \theta \right) \Gamma (g+1,\theta )}\right) ^{\frac{2}{D}}
\]

\[
\theta \equiv g+1+\frac{D}{2}
\]
$\Gamma (\nu ,z)$ is the complementary incomplete gamma function, $%
i\equiv I/I_{c}$ is the dimensionless electric current,
$I_{c}=\alpha \left( \left( g+1\right)
/G\overline{A}\right)^{D/2}$ is the critical current of the
transition into the resistive state, $\alpha $ is a form factor,
and $D$ is the fractal dimension of the cluster boundary. The
fractal dimension determines the scaling relation between
perimeter and area:
\begin{equation}
P^{1/D}\propto A^{1/2}  \label{eq4}
\end{equation}
where $P$ is the perimeter of a cluster of area $A$. Relation of Eq.~(%
\ref{eq4}) is consistent with the generalized Euclid theorem
according to which the ratios of the corresponding geometric
measures are equal when reduced to the same dimension \cite{man}.

In the particular case of $g=0$ the gamma-distribution of Eq.~(\ref{eq2})
reduces to the exponential one, $w\left( a\right) =\exp \left( -a\right) $,
for which $\overline{a}=\sigma _{a}=1$. This distribution of areas of the
normal phase clusters with fractal boundaries was realized in YBCO-based
film structures \cite{pla}, \cite{prb}.

The function of the critical current distribution of
Eq.~(\ref{eq3}) allows us to fully describe the effect of the
transport current on the trapped magnetic flux. Thus, the density
$n$ of vortices broken away from the pinning centers by the
current $i$ can be found:
\begin{equation}
n\left( i\right) =\frac{B}{\Phi _{0}}\int\limits_{0}^{i}f\left( i^{\prime
}\right) di^{\prime }=\frac{B}{\Phi _{0}}\frac{\Gamma \left(
g+1,Gi^{-2/D}\right) }{\Gamma \left( g+1\right) }  \label{eq5}
\end{equation}
where $B$ is the magnetic field, $\Phi _{0}\equiv hc/\left( 2e\right) $ is
the magnetic flux quantum, $h$ is the Planck constant, $c$ is the speed of
light, and $e$ is the electron charge. The integral in the right-hand part
of Eq.~(\ref{eq5}) is the cumulative probability function, which is a
measure of the number of clusters of critical current smaller than the
preset value of $i$.

\begin{figure}[tbp]
\epsfbox{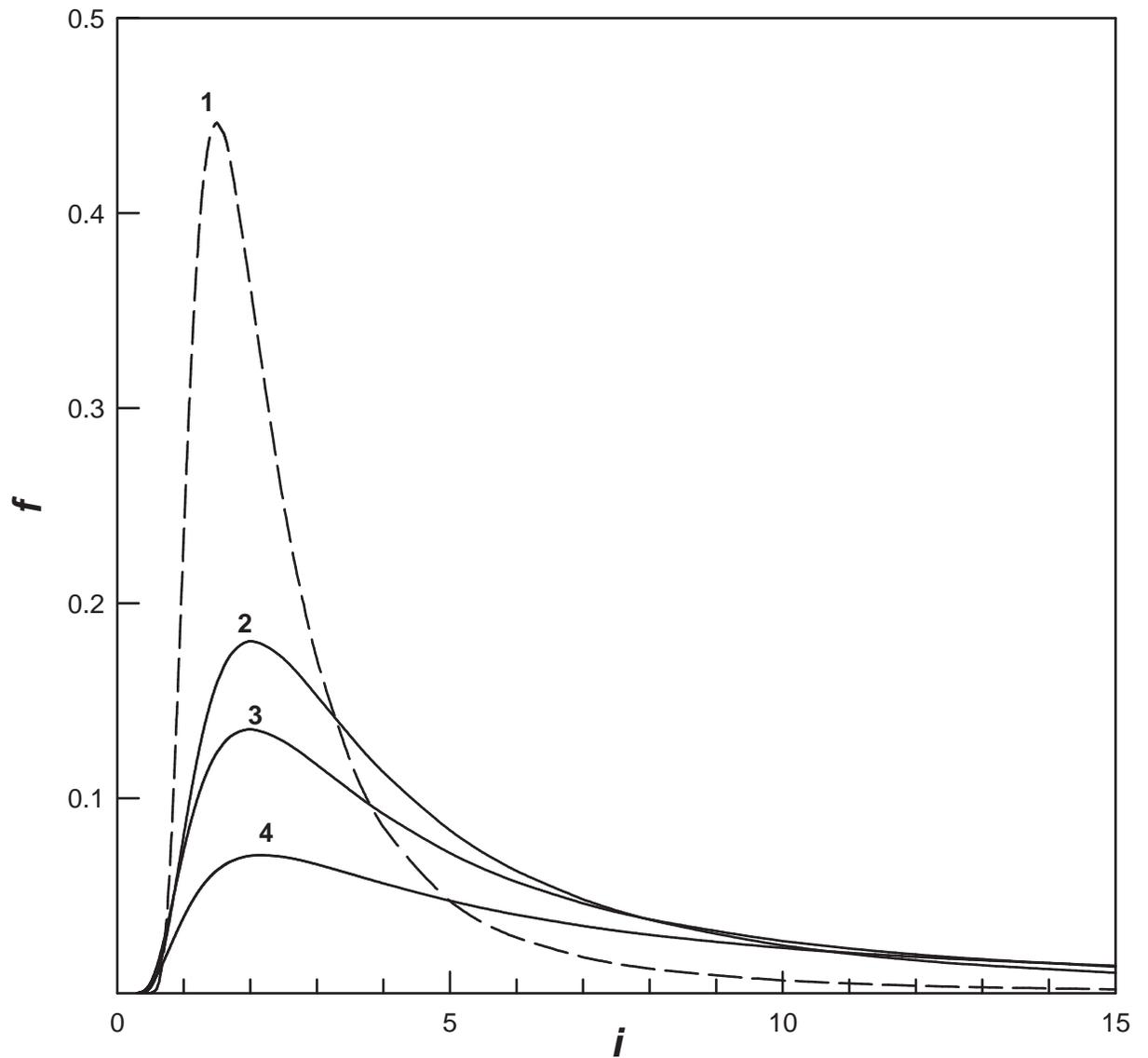}
\caption{Influence of the fractal dimension
$D$ and $g$-parameter of gamma-distribution on the distribution of
critical currents. The dotted line is drawn for the Euclidean
clusters ($D=1$); and the solid lines for the clusters with the
most fractal boundaries ($D=2$). Curve (1) corresponds to
the case of $D=1$, $g=0$; curve (2) of $D=2$, $g=0.5$; curve (3) of $D=2$, $%
g=0$; curve (4) of $D=2$, $g=-0.5$.}
\label{figure1}
\end{figure}

Figure \ref{figure1} shows how the fractal dimension of cluster
boundaries affects the critical current distribution. A comparison
between curves (1) and (3) shows that an increase in the fractal
dimension leads to a significant extension of the ``tail'' of the
distribution $f=f\left(
i\right) $. This effect is more pronounced for smaller values of $g$%
-parameter of gamma-distribution.

In the interval of currents $i>1$ the magnetic flux motion causes
the electric voltage across the sample, which goes into the
resistive state. Because of the finite resistance any current flow
is accompanied now by the energy dissipation. As for any hard
superconductor (type-II, with pinning centers) the presence of
dissipation in the resistive state does not necessarily implies
the destruction of the phase coherence. The superconducting state
collapses only when the dissipation exhibits an avalanche-like
growth due to the thermo-magnetic instability development.

In the resistive state the superconductor is adequately described
by the current-voltage characteristic. Using the fractal
distribution of critical currents of Eq.~(\ref{eq3}), we can find
the electric field arising from the magnetic flux motion after the
vortices have been broken away from the pinning centers. Since
each cluster of the normal phase contributes to the overall
distribution of critical currents, the voltage across the
superconductor $V=V\left( i\right) $ is an integral response to
the sum of contributions from all clusters:
\begin{equation}
V=R_{f}\int\limits_{0}^{i}\left( i-i^{\prime }\right) f\left( i^{\prime
}\right) di^{\prime }  \label{eq6}
\end{equation}
where $R_{f}$ is the flux flow resistance. The similar
representation for the voltage across the sample is frequently
used in the description of pinning of the vortex filament bundles
in superconductors \cite{war} as well as in the analysis of
critical scaling of the current-voltage characteristics \cite
{bro}, that is to say, in all cases where there is a distribution
of the depinning currents. In the present consideration we will
concentrate on the conclusions following from the properties of
fractal distribution of Eq.~(\ref{eq3}), so we  will not consider
any questions related to the possible dependence of the flux flow
resistance $R_{f}$ on the transport current.

After substitution of the distribution function of Eq.~(\ref{eq3})
into the convolution integral of Eq.~(\ref{eq6}) and carrying out
the integration, we eventually obtain an expression for the
voltage across the sample
\begin{equation}
\frac{V}{R_{f}}=\frac{1}{\Gamma \left( g+1\right) }\left( i\Gamma \left(
g+1,Gi^{-2/D}\right) -G^{D/2}\Gamma \left( g+1-\frac{D}{2},Gi^{-2/D}\right)
\right)  \label{eq7}
\end{equation}

\begin{figure}[tbp]
\epsfbox{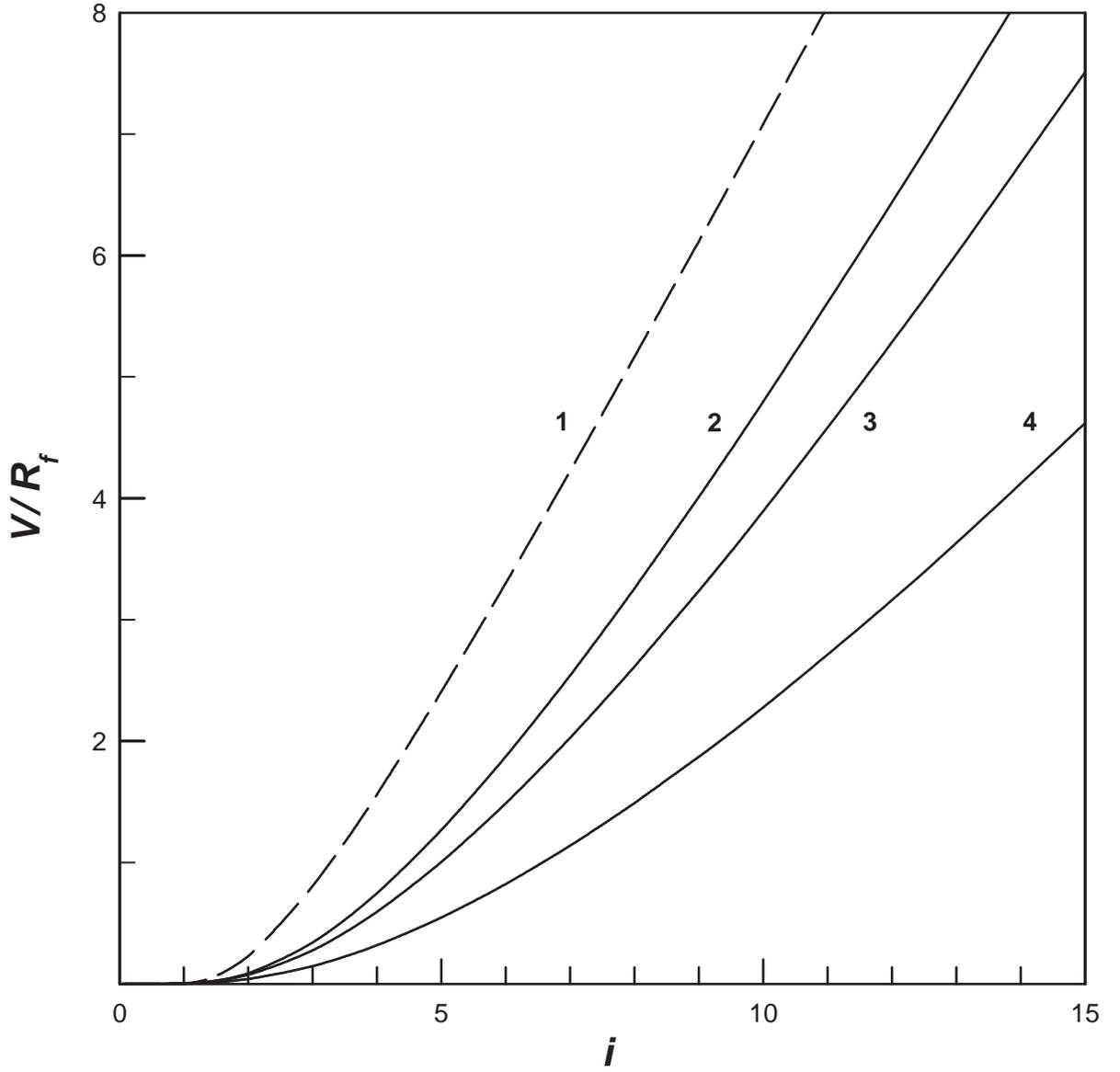}
\caption{The current--voltage characteristics at different values of $g$%
-parameter for Euclidean clusters ($D=1$, dotted line), and for
the clusters of the most fractal boundaries ($D=2$, solid line).
Curve (1) corresponds to the case of $D=1$, $g=0$; curve (2) of
$D=2$, $g=0.5$; curve (3) of $D=2$, $g=0$; curve (4) of $D=2$,
$g=-0.5 $.}
\label{figure2}
\end{figure}

Figure \ref{figure2} shows the current-voltage characteristics
calculated by formula of Eq.~(\ref{eq7}) for a superconductor with
fractal clusters of a normal phase. It should be noted that
similar current-voltage characteristics were observed in the
experiments on the measurement of the dynamic resistance of BPSCCO
composites containing silver inclusions \cite{pre}. A significant
drop in the voltage across a sample for all fractal dimensions is
observed beginning with the transport current $i=1$, which
coincides with the current of the transition into the resistive
state found in Ref.~\cite{pla} from the cumulative probability
function of the critical current distribution. For smaller
currents the trapped flux remains virtually unchanged because
there are no clusters with such a small depinning current.

As may be seen from Fig.~\ref{figure2}, the fractality of clusters
significantly reduces electric field arising from the magnetic
flux motion in a superconductor, and this effect is enhanced with
decreasing the $g$-parameter. This feature can be explained by
peculiarity of the exponential-hyperbolic distribution of
Eq.~(\ref{eq3}). An increase in the fractal dimension leads to
expansion of the critical current distribution toward greater
values of the current. At the same time, the total area under the
curve $f=f\left( i\right) $ remains unchanged because the
probability function of Eq.~(\ref {eq3}) is normalized to unity in
the whole interval of possible positive critical currents. This
implies that an increasing number of clusters, which can best trap
the magnetic flux, are involved into the process. Therefore, the
number of vortices broken away from the pinning centers by the
Lorentz force tends to decrease, so a smaller part of the flux can
move and, hence, an electric field of lower magnitude is
generated. The lower the electric field, the smaller energy is
dissipated as a result of the transport current passage, so a
decrease in the heat-evolution, which could induce the transition
into the normal state, leads to an increase in the critical
current for the superconductor containing such fractal clusters.
The maximum current-carrying capability is achieved upon
decreasing the $g$-parameter: in this case the clusters of small
size, which have the maximum critical currents, make the most
contribution to the overall distribution.

The probability re-distribution resulting from a change in the
fractal dimension is characterized by the variance of critical
currents and can be represented by the standard deviation
\[
\sigma _{i}=G^{\frac{D}{2}}\sqrt{\frac{\Gamma \left( g+1-D\right) }{\Gamma
\left( g+1\right) }-\left( \frac{\Gamma \left( g+1-D/2\right) }{\Gamma
\left( g+1\right) }\right) ^{2}}
\]

\begin{figure}[tbp]
\epsfbox{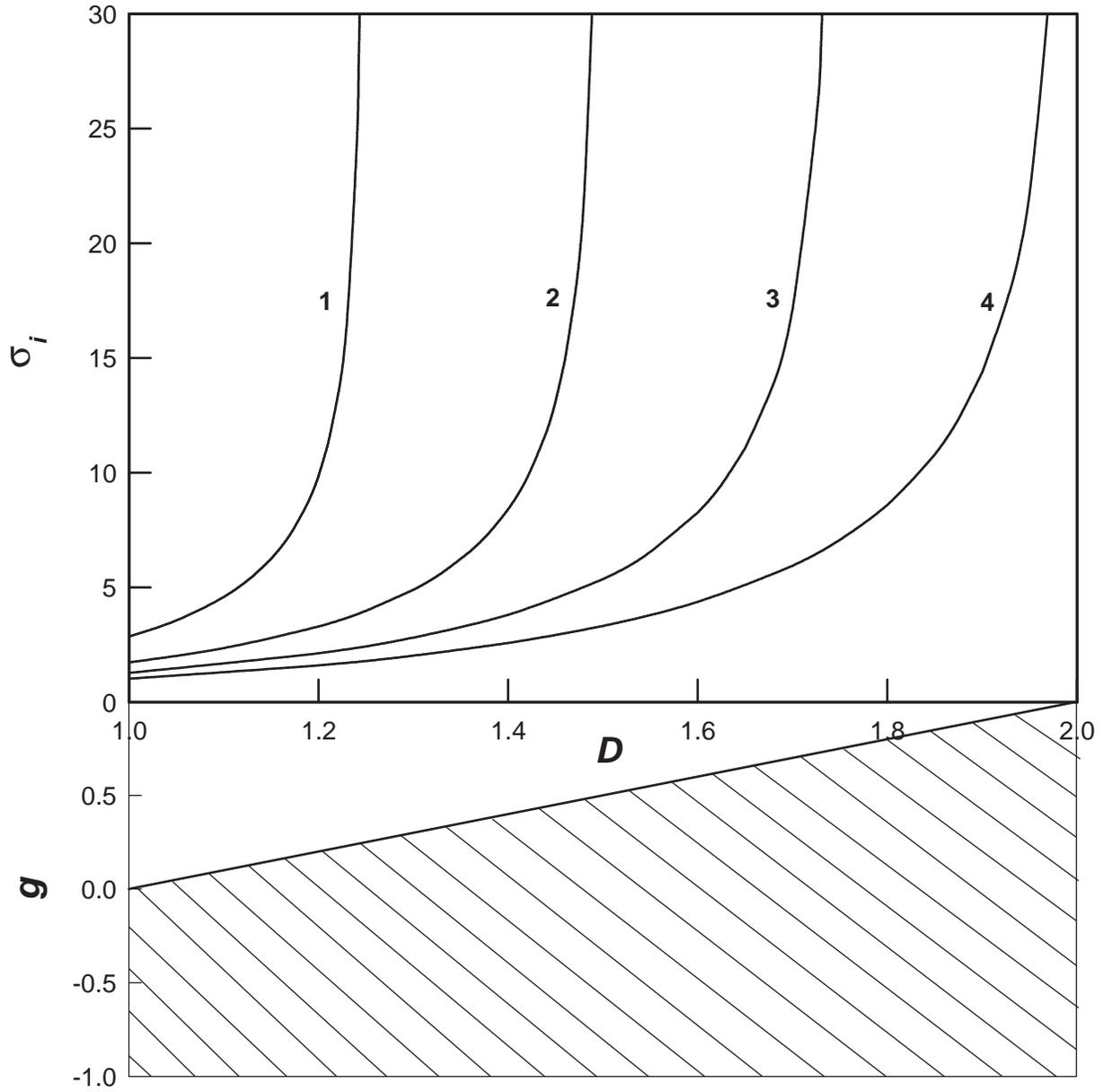}
\caption{Influence of the fractal dimension
of the cluster boundary on the
standard deviation of critical currents at different values of $g$%
-parameter. Curve (1) corresponds to the case of $g=0.25$; curve (2) of $%
g=0.5$; curve (3) of $g=0.75$; curve (4) of $g=1$. The region of giant
dispersion of the critical current ($D\geqslant g+1$) is shown by hatching.}
\label{figure3}
\end{figure}

Dependence of the critical current standard deviation on the fractal
dimension has a pronounced superlinear character (Fig.~\ref{figure3}). For $%
g<1$ there exists an interval of $D\geqslant g+1$, where the
dispersion of critical currents exhibits infinite growth. This
giant dispersion region is indicated by hatching in
Fig.~\ref{figure3}. The distributions with infinite variance are
well known in the probability theory, a classical example is the
Cauchy distribution \cite{hud}. However, such a behavior of the
exponential-hyperbolic distribution of Eq.~(\ref{eq3}) is of
special interest, since an increase in the variance leads to the
increase in critical currents. A statistical distribution with the
giant dispersion possesses an extremely extended ``tail''
corresponding to the contributions from clusters of high depinning
currents. To the present, a minimum value of $g$-parameter (equal
to zero) is realized in the YBCO-based composites with exponential
distribution of the areas of normal phase clusters \cite{pss}. The
films containing such clusters exhibit increased current-carrying
capacity. We may expect that superconductors containing the normal
phase clusters with the distribution of areas characterized by the
values of $g\leqslant D-1$ would provide for an additional
increase in the critical currents.

Thus, the fractal properties of normal phase clusters
significantly affect the dynamics of magnetic flux trapped in a
superconductor. This phenomenon is related to a radical change in
the distribution of critical currents caused by an increase in the
fractal dimension of cluster boundaries. The most important result
is that the fractality of the normal phase clusters enhances
pinning, thus preventing the destruction of superconductivity by
the transport current. This effect opens principally new
possibilities for increasing the critical currents in composite
superconductors by optimization of their geometric-morphological
characteristics.

\begin{center}
${\bf Acknowledgements\smallskip }$
\end{center}

This work is supported by the Russian Foundation for Basic Researches (Grant
No 02-02-17667).

\newpage

\newpage

\newpage

\end{document}